\documentclass[letterpaper,twocolumn,10pt]{article}
\usepackage{usenix2019,epsfig,endnotes}

\usepackage{amsmath,amssymb,amsfonts}
\usepackage{algorithmic}
\usepackage{graphicx}
\usepackage{textcomp}
\usepackage{appendix}

\usepackage{balance}  
\usepackage{graphics} 
\usepackage[T1]{fontenc}
\usepackage{txfonts}
\usepackage{mathptmx}
\usepackage{color}
\usepackage{booktabs}
\usepackage{textcomp}
\usepackage{microtype} 
\usepackage{ccicons}  
\usepackage{tabularx}
\usepackage{times}    
\usepackage{url}      
\usepackage{ulem} 
\usepackage{verbatim} 
\usepackage{multirow}
\usepackage{comment}
\usepackage{subfig}

\usepackage{amssymb}
\usepackage{mathrsfs}
\usepackage{dsfont}

\newcommand{\textred}[1]{\textcolor{red}{#1}}

\newcommand{\textmagenta}[1]{\textcolor{magenta}{#1}}
\ifx\noeditingmarks\undefined
   \newcommand{\pgwrapper}[2]{\textred{#1 #2}}
\else
   \newcommand{\pgwrapper}[2]{}
\fi
\ifx\noeditingmarks\undefined
\newcommand{\pgwrapperpurp}[2]{\textmagenta{#1 #2}}
\else
\newcommand{\pgwrapperpurp}[2]{}
\fi

\def\BibTeX{{\rm B\kern-.05em{\sc i\kern-.025em b}\kern-.08em
    T\kern-.1667em\lower.7ex\hbox{E}\kern-.125emX}}

\begin{document}

\date{}

\title{Quantifying the Security of Recognition Passwords: Gestures and Signatures}

\author{
{\rm Can Liu, Shridatt Sugrim, Gradeigh D. Clark, Janne Lindqvist}\\
Rutgers University
} 

\maketitle

\subsection*{Abstract}
Gesture and signature passwords are two-dimensional figures created by drawing on the surface of a touchscreen with one or more fingers. Prior results about their security  have used resilience to either shoulder surfing, a human observation attack, or dictionary attacks. These evaluations restrict generalizability since the results are: non-comparable to other password systems (e.g. PINs), harder to reproduce, and attacker-dependent.  Strong statements about the security of a password system use an analysis of the statistical distribution of the password space, which models a best-case attacker who guesses passwords in order of most likely to least likely. 

Estimating the distribution of recognition passwords is challenging because many different trials need to map to one password. In this paper, we solve this difficult problem by: (1) representing a recognition password of continuous data as a discrete alphabet set, and (2) estimating the password distribution through modeling  the unseen passwords. We use Symbolic Aggregate approXimation (SAX) to represent time series data as symbols and develop Markov chains to model recognition passwords. We use a partial guessing metric, which demonstrates how many guesses an attacker needs to crack a percentage of the entire space, to compare the security of the distributions for gestures, signatures, and Android unlock patterns. We found the lower bounds of the partial guessing metric of gestures and signatures are much higher than the upper bound of the partial guessing metric of Android unlock patterns.

\section{Introduction}

Passwords have been the dominant method for controlling access to a computing terminal since the 1960s~\cite{PasswordSecurity} and have survived despite a plethora of attempts to replace them with methods based on tokens or biometrics~\cite{Bonneau_CompareMetric}. The mobile computing era saw a shift among various types of user secrets~\cite{gestureMagzine}: text passwords remain the most popular technique on desktops~\cite{GuessAgain,realWorldMetrics,textNeuralNetworks}, whereas PINs and the $3\times3$ Android pattern unlock are the most popular secret types on mobile phones. The rise of fingerprint sensors has not blunted the momentum of passwords since one of the above methods is often used~\cite{GuessAgain,realWorldMetrics,textNeuralNetworks} as a fallback authentication method in the event of hardware failure, multiple unsuccessful unlocks, or device start-up.

\begin{figure}[!t]
	\centering
	\includegraphics[width=1\columnwidth]{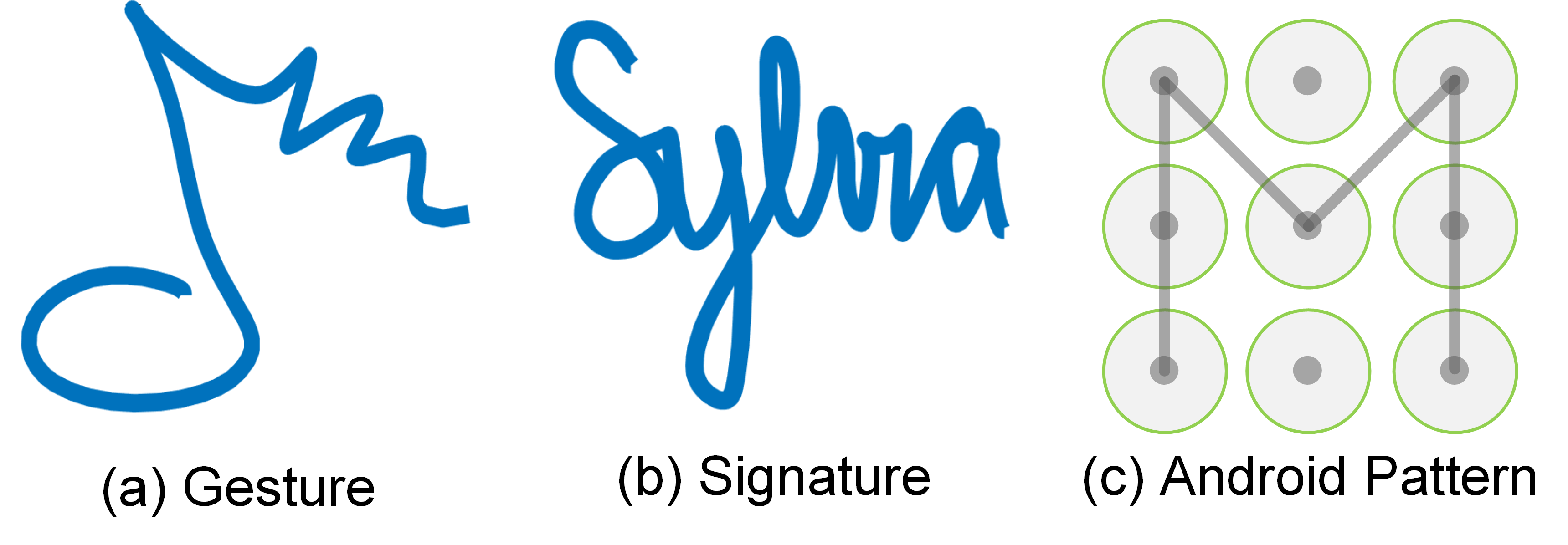}
	\caption{The examples matching and recognition passwords types. Recognition 
	passwords are gestures (a) and signatures (b), whereas matching password is Android patterns (c).}
	\label{fig:Figure_PasswordsExample}
\end{figure}

Despite differences between these three popular methods for the generation of user-chosen passwords -- patterns are graphical and PINs use a limited 0-9 character set -- each method is an example of a \textit{matching password}, which we define as a password that can be compared directly to a preselected pattern to determine a correct entry. Mobile computing has enabled the rise of gestures and signatures, two-dimensional figures that are drawn on the surface of a touchscreen with one or more fingers. We refer to gestures and signatures as \textit{recognition passwords} since they require a recognition algorithm to output a numeric measure of how correct a password attempt is. The recognition algorithm decides correctness of entry based on what figure was drawn and how it was drawn (e.g. acceleration of the finger through curves, timing between punctuations). Figure~\ref{fig:Figure_PasswordsExample} shows example passwords for gestures, signatures, and patterns.

Security is one of the crucial factors that determines the value of authentication 
systems. However, the advantages of novel recognition passwords are usually 
qualitatively evaluated based on shoulder surfing 
attacks~\cite{free_form_gesture, smudgeAttack,midair,AirAuth, kinwrite} or are 
quantitatively evaluated using dictionary 
attacks~\cite{text_PCFG,text_MarkovModel,JohnTheRipper,Hashcat,PasswordsPro}. 
Shoulder surfing evaluations are of limited utility because shoulder surfing 
attacks cannot be deployed at scale. The current evaluation of dictionary attacks 
focuses on a particular attacker. The efficiency of the attacker depends on 
several factors that weaken the generalization of experimental results, 
including the attacking strategy, dictionary choices, target passwords, and the 
size of the dictionaries and other datasets used. The results of these security evaluations are 
non-comparable since they depend highly on the attackers' behavior, which is not stable across experiments. 

Partial guessing metrics~\cite{partialGuess}, on the other hand, consider an 
idealized attacker that chooses candidate passwords based on a global 
knowledge of the distribution on user-chosen passwords for a system. These 
metrics quantitatively evaluate password security based on this distribution. 
They have been used to evaluate the security of matching 
passwords like Android pattern 
unlocks~\cite{android_pattern,AndroidStrengthMeter} and 
text passwords~\cite{partialGuess}. Partial guessing metrics enable a security comparison between different authentication methods because they rely on 
the size of the password space and the user preference for selecting passwords.

\begin{figure}[!t]
	\centering
	\includegraphics[width=0.8\columnwidth]{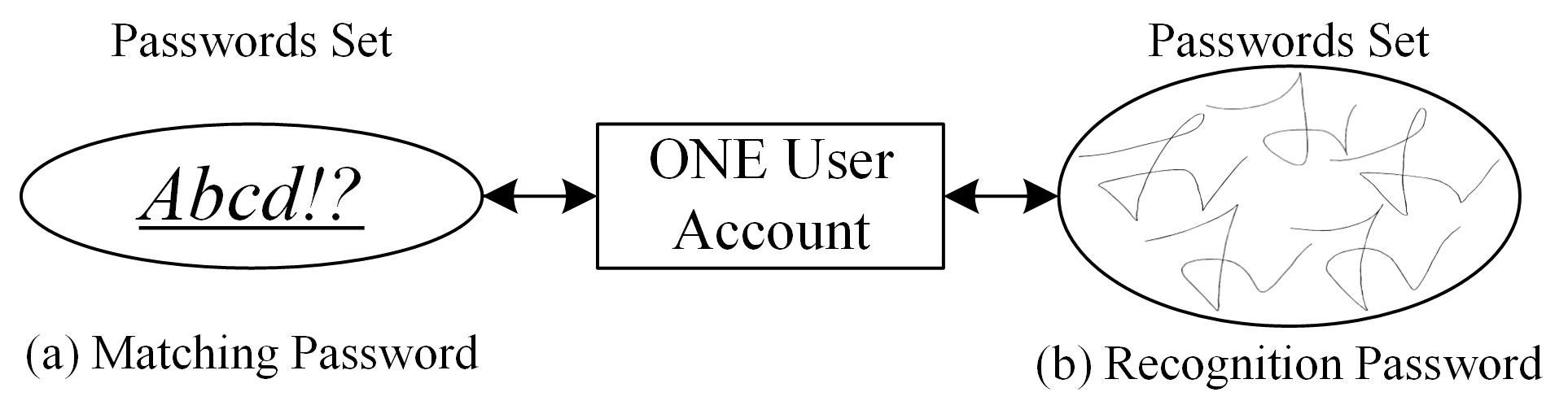}
	\caption{The many-to-one problem for enumerating the password space for recognition passwords. 
		A matching password has only one acceptable value, but  many different attempts at a 
		single recognition password that can map to a single user account. An attacker should efficiently 
		group the universe of password trials to enumerate the passwords distribution.}
	\label{fig:Figure_manyToOne}
\end{figure}

The challenge inherent to adapting a partial guessing metric for recognition 
passwords lies in the enumeration of the password space. For matching 
passwords, any change in a password results in a different password. For 
example, for text passwords $abc?$ and $abc!$ are not the same because they do 
not match. A small change in a recognition password, however, could be one of 
two things: a different password or an attempt at entering the known password. 
This is a many-to-one matching problem, and Figure~\ref{fig:Figure_manyToOne} 
shows these differences. The factors that affect the matching results of 
recognition passwords are 1) the similarity measurement between passwords and 
2) the threshold of similarity under which two slightly different passwords are 
regarded as the same. Both of these factors are chosen by the system designer and 
have limited influence on users' password selection preferences.

To approximate the number of possible passwords a many-to-one recognition 
password system has, we need an approximation method that can safely ignore the 
difference between slightly different password attempts from the same user. 
This approximation, however, needs to ensure that the differences between 
different users' passwords remain tangible. Our method was to construct an 
approximate recognizer that we can then use to compute the probabilities of 
password being chosen. We used the distinguishability 
between different users' passwords to show the validity of the approximation 
method. If the false positive and false negative rates of the approximate 
recognizer are comparable to actual recognizers, then the probabilities 
computed will be a good approximation to the true probabilities.

We used Symbolic Aggregate approXimation (SAX)~\cite{SAX}, a method that 
approximates time-series data through a short sequence of symbols, as the first 
step in the approximate password recognizer. Representing the password space as 
a sequence of symbols allows us to group together passwords that should be 
considered the same in a simple way. The recognizer compares the symbolic 
representation of two inputs rather than outputting a measure of similarity 
between two inputs. SAX enables us to convert a recognition password into a set 
of symbols and to analyze recognition passwords in a more manageable and less 
sparse password space.

\begin{figure}[!t]
	\centering
	\includegraphics[width=0.8\columnwidth]{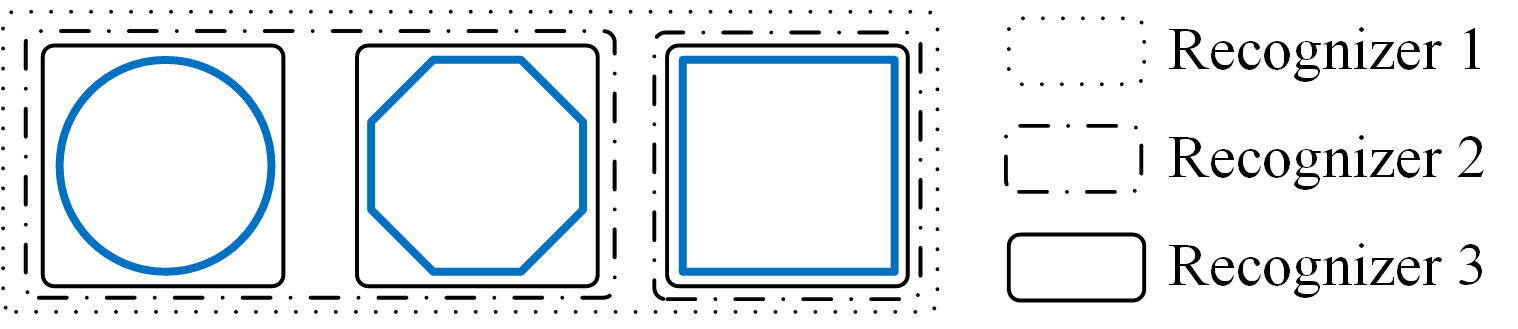}
	\caption{To motivate why our approach establishes a baseline for security 
	consider  a 
		password space of only three possible passwords. A coarse recognizer 
		(Recognizer 1) would 
		classify all shapes as the same password. A fine recognizer (Recognizer 
		3) identifies each shape as  different passwords, and thus increases 
		the size of the password space. Recognizer 2, represents the typical 
		case, similar shapes are identified as the same password. The number of 
		passwords depends on how similarity is measured. Our proposed method 
		will undercount the number of passwords because its many to one 
		mapping 
		makes it very coarse.}
	\label{fig:Figure_recognizerLowerBound}
\end{figure}

With SAX, our analysis of recognition passwords provides a baseline for the 
security of the true password distribution. The size of the password space is a 
function of recognizer granularity -- a finer recognizer will separate 
passwords with slight differences, and thus have a larger password space. 
To reduce the complexity of the password space,  we group similar 
passwords. This approach inherently undercounts the password space. In this 
way, we are providing a baseline. This is depicted in 
Figure~\ref{fig:Figure_recognizerLowerBound}.

Another challenge in applying a partial guessing metric to all novel 
authentication methods is the task of  collecting a large enough password 
dataset to reflect the distribution of passwords. We note that it is necessary 
to estimate the security of a novel authentication method with a small password 
dataset before it is widely deployed. Thus, we trained a Markov chain on 
the largest existing recognition password datasets. The Markov chain models the 
probabilities of transitions from one symbol to another. With the transition 
probabilities from the Markov chain, we enumerate and estimate the 
probabilities of all possible passwords in this space.

\textbf{In this paper, we present the first successful attempt to quantify the security of user-chosen secrets for recognition passwords and compare their security to that of matching passwords based on a realistic model of attacker behavior: a partial guessing metric.} We chose gestures and signatures as two examples of recognition passwords since their password distributions are affected by two different factors: 1) gesture passwords are mainly affected by the shape of the passwords that users select, and 2) signature passwords are mainly affected by the behavior that users employ to perform their passwords. Additionally, plenty of researchers have studied these two kinds of recognition passwords and made datasets available~\cite{free_form_gesture,ESM,Policy,Garda,guessingAttack,kinwrite,midair,AirAuth,AuthKinect,Biometric_rich,susig,mcyt,svc2004,gradeighpervasivegesture}.
We chose the Android unlock pattern as a representative example of a matching password system since prior studies have evaluated its security with partial guessing metrics~\cite{android_pattern,AndroidStrengthMeter}.

\textbf{We solve the many-to-one mapping problem for recognition passwords by discretizing them.} We applied SAX~\cite{SAX} to represent the time series data of recognition passwords as discrete 
symbols. By comparing the recognition performance of SAX to the commonly used recognition 
methods (Protractor~\cite{free_form_gesture,ESM,Policy}, 
DTW~\cite{guessingAttack,kinwrite,midair,AirAuth}, Garda~\cite{Garda}), we show that SAX retains enough information that separates discretized passwords to for them to remain distinguishable.

\textbf{We estimate upper and lower bounds of password security using a small 
dataset.} Specifically, we trained a Markov chain model with our largest 
existing password dataset and estimated the probabilities of all theoretical 
possible passwords. Since novel authentication methods always suffer from the 
common concern that a relatively small password dataset cannot cover the 
general password distribution, we used two strategies to deal with the 
uncovered passwords: 1) assigning them zero probability, and 2) assigning them 
a small probability value. These two strategies lead to the upper and lower 
bounds for the security estimation of passwords.

\textbf{Finally, we studied human selection bias in recognition passwords.} We found that people prefer to start recognition passwords in the upper left corner and end passwords in the bottom right corner on a 2-D touchscreen.

\section{Related Work}

Morris and Thompson~\cite{PasswordSecurity} were the first to document how user-chosen text passwords are vulnerable to dictionary attacks due to users choosing overlapping patterns with specific meanings rather than random text strings. Dictionary attack efficiency was improved by the development of probabilistic context-free grammar~\cite{text_PCFG}, John the Ripper~\cite{JohnTheRipper}, Hashcat~\cite{Hashcat}, Markov chains~\cite{text_MarkovModel}, and neural networks~\cite{textNeuralNetworks}. The key difference in these techniques is how the guessing attack is generated, but the idea remains the same:  humans are more likely to choose from a weak subspace instead of the full space of available passwords. 

Graphical passwords similarly have weak subspaces that allow dictionary attacks. Thorpe and Van Oorschot~\cite{Journal_DAS} found that people preferred to use symmetry to create passwords with Draw-A-Secret (DAS)~\cite{DAS}. Similarly, an analysis of Pass-Go~\cite{PassGo} showed that 49\% of users' passwords are alphanumeric or well-known symbols and that 40\% of users used either vertically or horizontally symmetric patterns to generate their passwords. Passfaces~\cite{passFaces} is a recognition-based variation of graphical passwords. A field study showed that people are more likely to use their own race's faces as passwords~\cite{Story}. Passpoints~\cite{PassPoints} is another cued-recall graphical password variation. Its major weaknesses are hotspots~\cite{PassPoints_Hotspots} (e.g.~image areas that people tend to choose) and patterns~\cite{PassPoints_Patterns} (e.g.~simple shapes that people are more likely to choose as PassPoints). All types of graphical passwords have a weak subspace that can be used for password cracking.

Attacks on gesture passwords have used shoulder surfing, in which attackers attempt to observe a user 
entering their password and then imitate it from memory~\cite{midair,AirAuth,kinwrite,free_form_gesture}. 
Shoulder surfing attacks cannot be deployed at scale and are not suitable for showing the general 
security of an authentication system except under targeted attacks. Guessing 
attacks~\cite{guessingAttack} were developed for gesture passwords and found that an efficient 
dictionary attack method could be derived from the fact that people prefer to use common graphical 
symbols as passwords. However, this method cannot crack any passwords that are not covered by the 
dictionary and cannot be used for analyzing a full password space. In this paper, we estimated the 
distribution of the entire password space with SAX and Markov chains and use partial guessing metric 
to model attacker behavior. This allows us to compare matching and recognition passwords.

Partial guessing has become an established metric for security analysis. Bonneau~\cite{partialGuess} 
proposed partial guessing metrics to model real-world attacks, wherein attackers only crack a portion 
of weak passwords and give up on guessing more difficult accounts. Uellenbeck et 
al.~\cite{android_pattern} applied partial guessing metric to evaluate the security of Android unlock 
patterns and PINs. However, they measured the guessing metric by cracking subsets of their 
collected data rather than through an estimation of the user-chosen distribution.
Song et al.~\cite{AndroidStrengthMeter} proposed a strength meter for Android unlock patterns and 
evaluated it using partial guessing metric. Aviv et al.~\cite{android_aviv} studied the impact on the 
security of Android unlock pattern when the grid size was increased from $3\times 3$ to $4\times4$ and 
found that this change does not improve the security of the Android unlock patterns. Kiesel et 
al.~\cite{mnemonicPassword} evaluated text password security using partial guessing metric for secure 
and mnemonic passwords. 

The research community has created various methods for recognizing gestures and signatures. Dynamic Time Warping (DTW) is the most widely used recognition method for gesture password systems~\cite{midair,kinwrite,AuthKinect,AirAuth} and is also widely used in signature authentication systems~\cite{authen_4,sigVQ_DTW}. Free-form gestures have used recognition methods based on cosine similarity~\cite{free_form_gesture,Policy,ESM}. Garda~\cite{Garda},  a multi-expert authentication system combining Gaussian Mixture Models and Protractor~\cite{protractor_method}, has been proposed for free-form gestures. Liu et al.~\cite{Garda} show that the recognition methods based on both cosine similarity and Garda are applicable for online signature passwords.

There are many studies on the guessability and security of matching passwords. However, to the best 
of our knowledge, there is no existing work on analyzing the security of recognition passwords because of their many-to-one problem for passwords. Here, we present the first attempt to quantitatively analyze matching passwords with partial guessing metric. We show that recognition passwords have a higher partial 
guessing metric than Android unlock patterns. We used SAX to encode the time sequence data of 
recognition passwords as a short sequence of discrete symbols. This outputs a recognition result comparable to the strongest freeform gesture recognizer -- Garda ~\cite{Garda}, which means 
that SAX is a reasonable discretization model for gesture passwords.

\section{Roadmap}

This section contains high-level descriptions of the logic involved in the later parts of the paper to help the reader better understand the presentation order of the paper and the meaning behind certain choices that were made. The technical details of our work follow immediately after this section.

We will present our ideas in the following order:

(1) How does an attacker behave to crack passwords?

(2) How to discretize the recognition password space?

(3) How to assign probability to a given password?

(4) How to estimate password security with a small dataset?

\subsection{Attacker Behavior: The Threat Model}
There are multiple ways an attacker can behave when targeting passwords. Their behavior can change based on the amount of information they have and their overall objective in trying to crack a specific password. An attacker could be focusing all their effort on trying to crack a single password without a thought to the many other passwords in the set. An attacker could use observations to gain information about the password, data mine a particular user to obtain ideas about what the password might contain, or attempt to steal the password through other means.

This targeting behavior is not useful for trying to evaluate the general security of an authentication method since it is not feasible for an attacker to exert this same level of effort for every individual user. Bonneau~\cite{partialGuess} outlined a framework for evaluating the security of text passwords based on the entire password distribution. In this framework, an attacker has access to a large number of accounts and is interested in maximizing the benefit of cracking a given account while minimizing the cost of cracking the same account. They do not have any prior information about the person whose password they are trying to crack. One might imagine that this is akin to obtaining a large list of email addresses and trying to crack the password for each one of those emails. \textbf{The attacker applies a number of guesses while trying to crack some percentage of the entire account set. By stopping after a fixed number of guesses, they do not waste resources on accounts with difficult passwords. By stopping after cracking some percentage of the targeted accounts, they will have achieved some minimal benefit.} We emphasize this point because it determines how we calculate security.

\begin{figure}[!t]
	\centering
	\includegraphics[width=1\columnwidth]{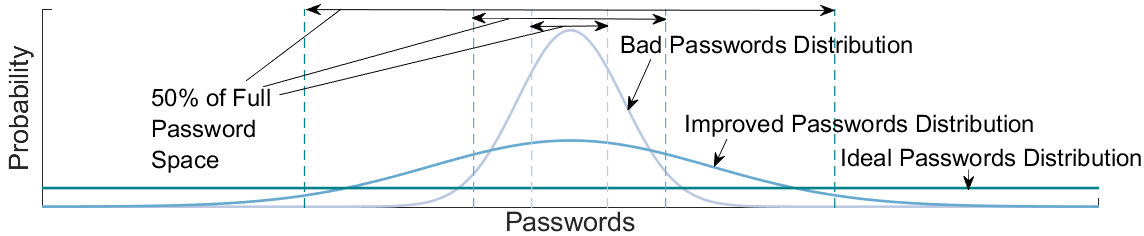}
	\caption{Example of different password distributions. The ideal password distribution is uniform, in which all passwords are equally probable. A weak password distribution usually has a narrow peak, meaning most user passwords fall in a small set: the weak set. An informed attacker would use the most probable passwords in the weak set when guessing. We can improve the passwords security by broadening the set that people are more likely to select from.}
	\label{fig:passwordDistribution}
\end{figure}

Secret selection by humans is usually biased, creating a subset of more likely passwords in the distribution that we refer to as the weak set of passwords. An attacker who is informed about the password distribution is capable of ordering their guesses from most likely to least likely. Therefore, password systems with a more concentrated distribution are more likely to be cracked and thus have lower security. Figure~\ref{fig:passwordDistribution} illustrates three types of password systems with different distributions. An ideal password system would have a uniform distribution, with all passwords being equally likely. This would prevent an attacker from gaining an advantage by ordering attacks, and the attacker would therefore have to guess randomly. \textbf{The security of a password system is tied to the distribution of user choice.}

Given this attacker behavior, the security metrics of biometric systems, like True Positive Rates and False Positive Rates, are misleading for assessing the security of recognition passwords. Typical biometric systems do not have a component of user choice. They are based on immutable human characteristics that are highly differentiable -- fingerprints, for example -- and cannot be changed at will. The primary security mechanism for biometrics is the high degree of separation among thousands of people. Therefore, it makes sense to discuss security with TPR and FPR for biometric systems.

For a recognition password, security is primarily derived from the choice of the secret-- that is, how likely it is that another person or attacker will select the same password. TPR and FPR measure the accuracy of a recognizer in separating out obviously different passwords; they are not a statement of security. The important metric here is the distribution of the user choice of passwords, and we therefore need an efficient recognizer to estimate that distribution.

\subsection{Research Challenge: Discretizing the Recognition Password Space}

Based on the above subsection, we now know how an intelligent attacker will behave towards a large number of accounts: attackers order their attacks efficiently in order not to waste effort on accounts that are too difficult to crack. The attacker should select the most likely passwords and move from there, and they should try different users' passwords instead of trying different variations of the same user's password. Both of these needs lead to one question: how can one transform all variations of a given password into a simple representation?

The main challenge here is that recognition passwords are a many-to-one mapping. In our paper, we perform this transformation using Symbolic Aggregate ApproXimation (SAX). SAX transforms time-series data into a symbolic set. Passwords that are similar to each other, when passed through SAX, become the same character string. In this way, it is easy to group the passwords together.

\subsection{Research Challenge: Enumerating and Assigning Probabilities to the Passwords}

By representing the long time series data of recognition passwords with short discrete SAX symbols, we are capable of enumerating the entire password distribution by listing all possible combinations of strings together. As such, it is possible to take the entire password set into account. However, assigning probabilities to these newly generated passwords is still an issue.

The representation of a password as a string has benefits besides solving the counting problem: it 
can be used in combination with Markov chains. The guiding principle behind a Markov chain is 
that the next symbol in a human-chosen string depends on some number of the previously 
chosen symbols. This logic comes from the intuition that, when given a partial example of a text 
string like $gestu...$, it is highly likely the whole string with the remaining characters is 
$gesture$.

The attacker transforms every recognition password in their corpus of prior data into a text string. Then, they pass these text strings through the Markov chain to obtain probabilities for every possible string in the set. This estimation method has been used in the past with text passwords~\cite{text_MarkovModel,firstMarkovPW} and Android unlock patterns~\cite{android_pattern,AndroidStrengthMeter}. Thus, the attacker can assign probabilities to every generated password to obtain a full distribution that they can deploy to attack a set of accounts.

\subsection{Research Challenge: Estimate Password Security with A Small Dataset}

\begin{figure}[!t]
	\centering
	\includegraphics[width=1\columnwidth]{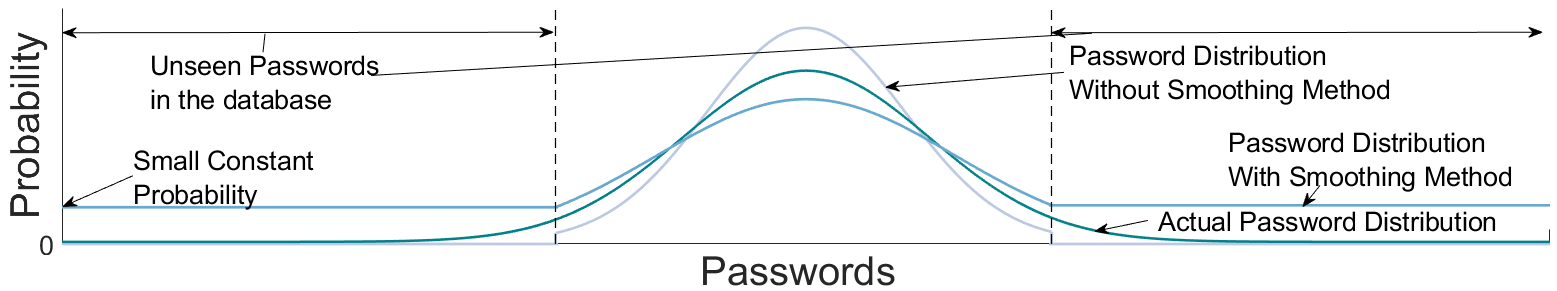}
	\caption{Example of passwords distributions with different methods for addressing the uncovered 
		passwords issue. The actual passwords distribution represents the real passwords distribution. The 
		distributions with and without smoothing methods represent the passwords distribution with the 
		two methods for addressing the uncovered passwords issue.}
	\label{fig:passwordDistributionSmooth}
\end{figure}

One of the key difficulties in generating Markov chain estimates of a password space distribution is completeness. A model is complete when it assigns a non-zero probability to all possible passwords. If a model is incomplete, then some passwords have zero probability -- we call these \textit{uncovered} passwords. When the completeness of a model is poor, the estimated password distribution will skip over passwords that are likely to be selected by people but are not covered by the trained Markov chain model.

There are two factors that may lead passwords to remain uncovered. First, the passwords may be very unlikely to be selected by people. Second, the passwords may be likely to be selected by people, but the dataset may not cover them. Based on these two reasons, we used two strategies to deal with the covered passwords: (1) leaving the uncovered passwords as zero probability, and
(2) assigning small fixed probabilities to the uncovered passwords. The first method can help us eliminate impossible passwords, and the second method avoids a situation in which potential passwords are skipped. Figure~\ref{fig:passwordDistributionSmooth} illustrates the password distribution that results after each of these two methods is applied. We found that the password distribution that results from the application of a smoothing method has a peak wider than the actual password distribution while the distribution that appears without smoothing has a peak narrower than the actual password distribution. This means the model of password distribution that does not use a smoothing method underestimates password security and the model that does use a smoothing method overestimates security. Therefore, these two models will provide upper and lower bounds on the password security estimation.

\section{Password Datasets}

We aggregated the largest available gesture dataset and a large signature dataset to analyze recognition passwords. We used gesture datasets from the following studies: FreeForm~\cite{free_form_gesture}, Wild~\cite{ESM}, and GuessAttack~\cite{guessingAttack}. In each of these studies, the participants are asked to create accounts with gestures as passwords. The gesture passwords for each account need to be replicated at least once. In total, there are 2656 gesture password samples from 655 types of gesture passwords. The signatures dataset includes three publicly available datasets: SUSig~\cite{susig}, MCYT-100~\cite{mcyt}, and SVC2004~\cite{svc2004}. There are 5180 signature samples in total from 234 participants, and all of them were collected in a laboratory setting. Table~\ref{tab:Datasets} summarizes the datasets.

To analyze Android unlock patterns as a reference for matching passwords, we requested and obtained 113 defensive Android unlock patterns and 573 offensive Android unlock patterns from 113 participants~\cite{android_pattern}. To generate defensive patterns, participants were asked to create Android unlock patterns that are difficult for others to crack. Similarly, to create an offensive pattern, participants created Android unlock patterns they felt were most likely to crack other users' patterns. Our study was approved by the Institutional Review Board (IRB) of Rutgers University.

\begin{table}[]\
\small
\centering
\begin{tabular}{cccc}
\hline
Dataset           & PW \# & Dataset  & PW \# \\ \hline
FreeForm~\cite{free_form_gesture}          & 684          & SUSig~\cite{susig}    & 1880         \\
Wild~\cite{ESM}              & 1536         & MCYT~\cite{mcyt} & 2500         \\
GuessAttack~\cite{guessingAttack}       & 436         & SVC2004~\cite{svc2004}  & 800          \\ \hline
Android (Def)~\cite{android_pattern} & 113         & Android (Off)~\cite{android_pattern} & 573      \\ \hline

\end{tabular}
\caption{Summary of analyzed datasets. The top left three datasets are gestures, the top right 
three are signatures. The bottom two are Android unlock patterns 
datasets.}
\label{tab:Datasets}
\end{table}

\section{Discretization of Recognition Passwords}

Before estimating the distribution of the password space, we must transform recognition passwords into a set of symbols. With a known set of symbols and sufficient training data, we can later use a Markov chain to generate the distribution of passwords we have not yet seen. 

In this section, we describe the process of representing recognition passwords with SAX. We discuss the many-to-one mapping of gestures to symbol sequences and how that mapping is used to populate the training data set. Finally, we discuss the method used to determine the parameters used in 2-D SAX.

\begin{figure*}[!t]
\centering
\includegraphics[width=2\columnwidth]{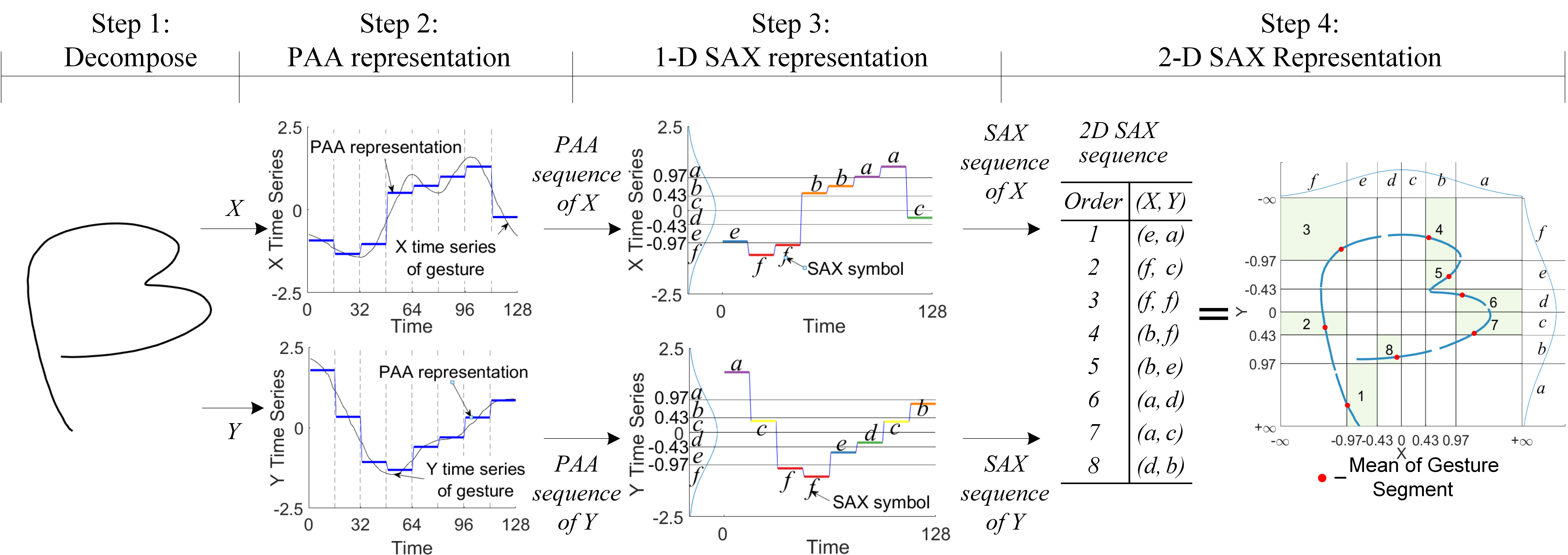}
\caption{Representing recognition passwords as a sequence of symbols using 2-D SAX. First, the 
password is decomposed into X and Y 1-D coordinate time sequences. In Step 2, each time 
sequence is normalized to have its mean set to zero and the standard deviation equal to one. 
The time sequence is then evenly segmented using PAA~\cite{PAA} into eight subsequences. 
SAX then maps the means of the eight subsequences into the six symbols: $a,b,c,d,e,f$. The 
boundaries of the six symbols are calculated using the distribution on the left, where each 
symbol is defined to have equal probability. In Step 3, we combine the SAX sequence of X and Y 
to form a 2-D SAX sequence, which can also be represented as a 2-D map as seen 
in the rightmost figure.}
\label{fig:Figure_SAX}
\end{figure*}

\subsection{Represent by 2-D SAX}

We used Symbolic Aggregate approXimation (SAX)~\cite{SAX_original} to normalize and discretize the 
time sequence data of a recognition password so that it can be represented as short sequence of 
symbols. SAX uses a sequence of symbols with a fixed length $\omega$ to represent a time series data 
of length $n$, where $\omega\ll n$~\cite{SAX_original}. Figure~\ref{fig:Figure_SAX} shows the steps for 
approximating a recognition password with 2-D SAX.

\begin{enumerate}
	\item Decompose password. We decomposed a gesture into two 1-D time sequences: X time series and Y time series. The time series of X and Y are normalized to be zero mean and unit standard deviation. 
	\item PAA representation. We approximate a password's 1-D time sequences by Piecewise Aggregate Approximation (PAA)~\cite{PAA}. PAA approximates a time series by segmenting it into $\omega$ equal-length subsequences and representing each subsequence by its mean. 

	\item 1-D SAX representation, we use SAX~\cite{SAX_original} to map the value of PAA representation into different symbols based on the partitioned ranges. Each value range is chosen to have the same probability according to the normalized distribution, which is an important step to guarantee that each symbols are chosen with the same probability.  In our example in Step 3 of Figure~\ref{fig:Figure_SAX}, the normal distribution is divided into six equal probabilities ranges with five boundaries, \{\textit{-0.97, -0.43, 0, 0.43, 0.97}\} and the six ranges are represented by six symbols, \{\textit{a,b,c,d,e,f}\}. The five boundary values are obtained by using the inverse-CDF of the standard normal distribution, which outputs the point at which a certain amount of probability is contained. If there are six equiprobable regions, then $CDF^{-1}(\frac{1}{6})=-0.97$, $CDF^{-1}(\frac{2}{6})=-0.43$, and so on. The point at which zero probability is contained is $-\infty$ and the point of all probability is $\infty$.
	\item 2-D SAX representation. 2-D SAX is simply a combination of SAX sequences of X and Y. It also can be represented by a 2-D matrix on the rightmost graph in Figure~\ref{fig:Figure_SAX} following the order of the cells. The rightmost graph shows the area of a gesture represented by 36 cells with different value combinations in X and Y coordinates and the gesture is chopped into eight pieces based on PAA. For each gesture piece, SAX used its means in X and Y coordinates to assign the 36 SAX cells.	
\end{enumerate}

We used and extended MINDIST function, which is defined for 1-D SAX~\cite{SAX}, to measure 
the similarity of 2-D SAX representations of recognition passwords. With the original MINDIST, we 
measure the similarity of two 1-D SAX sequences 
$\hat{Q}={\hat{q}_1,...,\hat{q}_\omega}$ and $\hat{C}={\hat{c}_1,...,\hat{c}_\omega}$ as following:

\begin{equation}
    MINDIST(\hat{Q},\hat{C})=\sqrt[n/\omega]{\sum_{i=1}^{\omega}(dist(\hat{q}_i,\hat{c}_i))}
\end{equation}

\begin{center}
$dist(\hat{q},\hat{c}) = \begin{cases} 0, & \mbox{if } |\hat{q}-\hat{c}|\leq 1 \\ \beta_{max(\hat{q},\hat{c})}-\beta_{min(\hat{q},\hat{c})}, & \mbox{otherwise} \end{cases}$
\end{center}

Where $n$ is the length of the original time sequence and $\omega$ is the length of the time sequence represented by SAX. $dist()$ is used to measure the distance between two symbols of SAX. $\beta$ is the set of boundaries of the symbols in SAX. The MINDIST codifies the requirements for two sequences to be regarded as distinct. 

The distance is straightforward to calculate once the boundaries are known. Assume there are six symbols: $a$, $b$, $c$, $d$, $e$, $f$ as seen in Figure~\ref{fig:Figure_SAX}. The five boundaries separating the six equiprobable ranges in the normalized distribution are \{\textit{0.97, 0.43, 0, -0.43, -0.97}\}, according to the inverse-CDF of the normal distribution. Since $|a-b|=1$, $dist(a,b)=0$ and $|a-c|=2$, $dist(a,c)=\beta_{max(a,c)}-\beta_{min(a,c)}=\beta_{a}-\beta_{c}=0.97-(0.43)=0.54$. The \textit{dist()} function can be implemented by a lookup table as shown in Table~\ref{distFunc}.

\begin{table}[]
	\small
\centering
\begin{tabular}{ccccccc}
                                & \textit{a}                & \textit{b}                & \textit{c}                & \textit{d}                & \textit{e}                & \textit{f}                \\ \cline{2-7} 
\multicolumn{1}{c|}{\textit{a}} & \multicolumn{1}{c|}{0}    & \multicolumn{1}{c|}{0}    & \multicolumn{1}{c|}{0.54} & \multicolumn{1}{c|}{0.97} & \multicolumn{1}{c|}{1.4}  & \multicolumn{1}{c|}{1.94} \\ \cline{2-7} 
\multicolumn{1}{c|}{\textit{b}} & \multicolumn{1}{c|}{0}    & \multicolumn{1}{c|}{0}    & \multicolumn{1}{c|}{0}    & \multicolumn{1}{c|}{0.43} & \multicolumn{1}{c|}{0.86} & \multicolumn{1}{c|}{1.4}  \\ \cline{2-7} 
\multicolumn{1}{c|}{\textit{c}} & \multicolumn{1}{c|}{0.54} & \multicolumn{1}{c|}{0}    & \multicolumn{1}{c|}{0}    & \multicolumn{1}{c|}{0}    & \multicolumn{1}{c|}{0.43} & \multicolumn{1}{c|}{0.97} \\ \cline{2-7} 
\multicolumn{1}{c|}{\textit{d}} & \multicolumn{1}{c|}{0.97} & \multicolumn{1}{c|}{0.43} & \multicolumn{1}{c|}{0}    & \multicolumn{1}{c|}{0}    & \multicolumn{1}{c|}{0}    & \multicolumn{1}{c|}{0.54} \\ \cline{2-7} 
\multicolumn{1}{c|}{\textit{e}} & \multicolumn{1}{c|}{1.4}  & \multicolumn{1}{c|}{0.86} & \multicolumn{1}{c|}{0.43} & \multicolumn{1}{c|}{0}    & \multicolumn{1}{c|}{0}    & \multicolumn{1}{c|}{0}    \\ \cline{2-7} 
\multicolumn{1}{c|}{\textit{f}} & \multicolumn{1}{c|}{1.94} & \multicolumn{1}{c|}{1.4}  & \multicolumn{1}{c|}{0.97} & \multicolumn{1}{c|}{0.54} & \multicolumn{1}{c|}{0}    & \multicolumn{1}{c|}{0}    \\ \cline{2-7} 
\end{tabular}
\caption{A lookup table for \textit{dist()} in MINDIST function when there are six possible symbols in SAX (i.e. $\beta=6$). The distance between two symbols can be easily found in the table. For example, $dist(d,f)=0.54$.}
\label{distFunc}
\end{table}

Previously, $\hat{Q}$ and $\hat{C}$ represent a time sequence of one dimension; in our modification, we define that $\hat{Q}$ and $\hat{C}$ represent a time sequence of $D$ dimensions. Accordingly, the distance between the $D$ dimensions is simply the sum of the distance of the one dimension symbols.

\subsection{Determining Parameters in 2-D SAX}

We have described how 1-D and 2-D symbols can be transformed into a symbolic sequence 
using SAX. However, there are important parameters that need to be determined rigorously in order to 
obtain the tightest bound possible on the password distribution. In order to discretize recognition 
passwords with 2-D SAX, we need to determine two parameters: the length of a symbolic 
sequence, $\omega$, and the alphabet of symbols, $\beta$. Increasing the values of the 
parameters increases the overall size of the password space -- by analogy, creating a new character for the Roman alphabet increases the space size for text passwords. 
The larger the parameters, the larger the difference between the different variations of the same 
password. Increasing the size of the parameters conflicts with the main goal of discretization, 
which is to narrow the difference between the variations of the same password. Thus we would 
like to minimize  $\omega$ and $\beta$. However, if we make the parameters too small, we will 
reach a point where no passwords are distinguishable. Thus we need to choose the smallest 
possible parameters that meet some minimum criteria of distinguish-ability (e.g. circles and 
squares are still seen to be different).
In this section, we will introduce Receiver Operating Characteristic~(ROC) curve and the Area Under Receiver Operating Characteristic curve~(AUROC) as evaluation metrics for $\omega$ and $\beta$. Then, we will present the optimal parameters for 2-D SAX.

\subsubsection{Receiver Operating Characteristic (ROC) curve}

The ROC curve is used to measure the recognition performance of recognition password systems. It is drawn by plotting the True Positive Rate and False Positive Rate with different recognition thresholds. The closer the curve is to the upper left corner, the better the recognition performance.

\begin{figure}[!t]
	\centering
	\includegraphics[width=1\columnwidth]{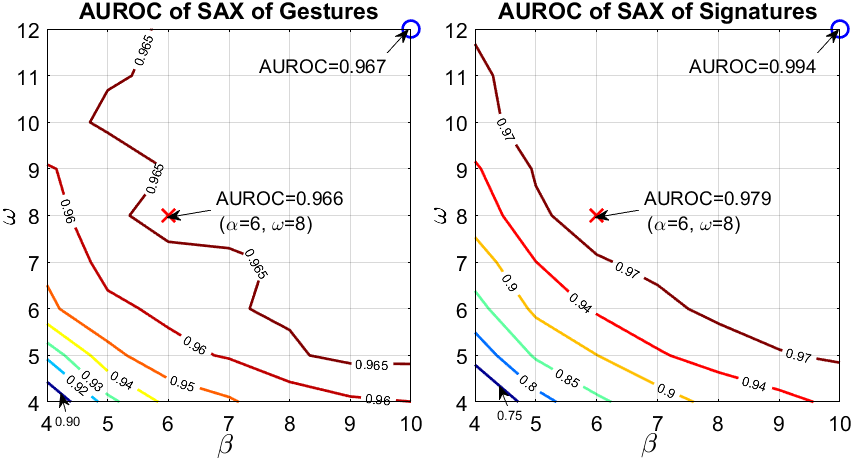}
	\caption{A contour plot showing isolines of AUROC for signature and gesture datasets based 
	on different values of $\omega$ and $\beta$. When $\omega \geq 8$ and $\beta \geq 6$, 
	the AUROC of SAX of gestures and signatures do not change by more than 0.001 and 0.015, 
	respectively. The isolines are erratic since the AUROC is highly dependent on the combination 
	of $\beta$ and $\omega$; these two parameters are not 
	orthogonal. We observed the that AUROC gradually decreases when $\omega$ and $\beta$ 
	increase.}
	\label{fig:Figure_AlphaOmega}
\end{figure}

\subsubsection{Area Under ROC~(AUROC) curve}

Because the recognition performance of different recognizers is evaluated on the same datasets, we can evaluate their distinguishability by computing the Area Under ROC (AUROC) curve. AUROC reflects the probability that a randomly chosen true password is ranked higher than a randomly chosen false password~\cite{hanley1982meaning}. It measures the distinguishability between the positive and negative samples. The higher of the AUROC, the more distinguishable between the samples.

\subsubsection{Optimal values of parameters $\omega$ and $\beta$}

Based on our analysis, we found that the length of a symbolic sequence, $\omega = 8$, and the alphabet of symbols, $\beta = 6$ balances the AUROC. Figure~\ref{fig:Figure_AlphaOmega} shows that when $\omega \geq 8$ and $\beta \geq 6$, the AUROC of the SAX recognizer for both gestures and signatures does not change significantly. This implies that when $\omega \geq 8$ and $\beta \geq 6$, the gestures and signatures from different users are only slightly more distinguishable.

\subsection{Recognition Performance of SAX}

\begin{figure}[!t]
	\centering
	\includegraphics[width=1\columnwidth]{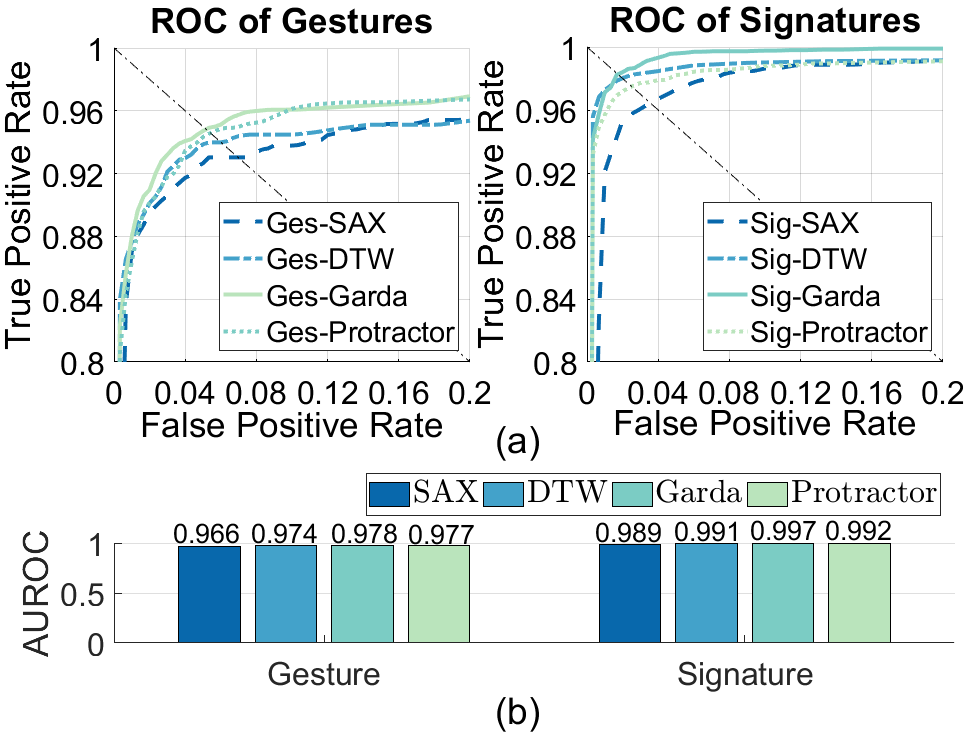}
	\caption{ROC curves and Area Under ROC (AUROC) of four recognizers for gestures and 
	signatures. From (a) ROC curves, we see SAX is only slightly worse than the other three 
	recognizers. From (b) AUROC, we observe that the AUROC of SAX is comparable to the other 
	recognizers. The takeaway is there is no large difference between the four recognizers.}
	\label{fig:Figure_ROC}
\end{figure}

SAX strips information away from recognition passwords through discretization. The rightmost side of Figure 6 also shows that gestures are significantly distorted by SAX. The distinguishability of  recognition passwords is a metric that shows whether SAX accurately models user behavior in drawing recognition passwords. To show that discretized passwords still reliably distinguish passwords among different users, we can examine the recognition performance of those discretized passwords with the ROC and AUROC. If the recognition performance of SAX is as good as that of other state-of-the-art recognizers, we can conclude that SAX is a valid discretizing method for recognition passwords.

The ROC curves for SAX have results comparable to those of other three recognizers. 
Figure~\ref{fig:Figure_ROC} (a) shows signature and gesture ROC curves for the four recognizers: 
SAX, Protractor~\cite{free_form_gesture,ESM,Policy}, 
DTW~\cite{guessingAttack,kinwrite,midair,MotionAuth,AirAuth}, and Garda~\cite{Garda}. Note 
that SAX is not meant to be the best recognizer -- our goal is to demonstrate that the symbolic 
representation maintains enough detail to distinguish gestures well enough. In this regard, the fact that it has performance that is comparable to other recognition methods allows us to consider SAX a success in this case.
We also evaluated SAX using the AUROC to determine the distinguishability between user and 
attacker passwords. Figure~\ref{fig:Figure_ROC} (b) shows the AUROC of four recognizers with 
gesture and signature datasets. We found that the AUROC values of SAX are close to those of the other three recognizers. This demonstrates that SAX has an ability to distinguish positive samples from negative ones that is comparable to those of the other recognizers.

In summary, representing recognition passwords with SAX can reduce the password space while maintaining the distinguishability of passwords.

\section{Recognition Password Distributions}

We have mapped the recognition passwords to a countable password space using SAX. We now need to estimate the user-chosen password distribution. Collecting data is not enough to do this, as it is very difficult to collect hundreds of thousands of passwords to represent the distribution. With SAX, we can enumerate the entire password space as a combination of all possible strings. A Markov chain will be trained with prior data to assign probabilities to the generated passwords. 

\subsection{Markov Chain}
Markov chains have been used to estimate the distribution for both text passwords~\cite{text_MarkovModel,firstMarkovPW} and Android unlock patterns~\cite{android_pattern,AndroidStrengthMeter}. The probability is computed as the product of conditional probabilities which represent the likelihood of transitioning from one symbol to the next in a sequence. These transition probabilities are estimated by their relative frequency of occurrence in a known data set. The guiding principle behind in a Markov chain is that the next symbol in a human-chosen string depends on some number of the previously chosen symbols. 

An \textit{n-gram} Markov chain predicts the next observation in a string based on the previous $n-1$ observations. To build Markov chains, there are two types of conditional probabilities that need to be estimated: 1) the probability of the starting symbol; 2) the probability of the transition to the next symbol given the previous $n-1$ symbols. For example, to build a 2-gram Markov chain for a sequence $s=\{s_1,s_2,...,s_8\}$, we need to estimate: 1) the starting symbol probability $p(s_1)$; 2) the transition probability $p(s_{i+1} | s_i)\;i \neq 0$. Then, we can calculate the probability of the sequence $s$ as $P(s)=P(s_1)P(s_2|s_1)...P(s_7|s_6)P(s_8|s_7)$.

To model the passwords based on an n-gram Markov chain, we need to determine n and our smoothing methods.

\subsubsection{Selecting $n$}

The value of $n$ defines both the number of previous symbols on which a transition depends as well as the length of the start sequence. Take the previous 2-D SAX sequences as an example: when $n=4$, the probability of a certain symbol appearing should be based on the prior three symbols. The start sequence should include various combinations of the first three symbols of a sequence, which makes for $36^3=46,656$ choices. A larger value of $n$ can yield more accurate predictions of the sequence by accounting for longer historical correlations. However, this gain in accuracy can also lead to over-fitting as many transitions may be assigned zero probability because the preceding sequence is never observed.

To determine a practical value of $n$ for the amount of data we have available, we can estimate the how large the zero probability blocks are by computing the expected number of times a given start sequence will be observed as a function of $n$. To simplify this problem, we assume that each start sequence is equally likely. Therefore, the expected number of observations of any particular starting sequence is as follows:

\begin{equation}
E(observation)=\frac{T}{(\beta^2)^{n-1}}=\frac{T}{(36)^{n-1}}
\end{equation}

$T$ is the total number of passwords in the data set. For the gesture dataset, $T=3245$, and for 
the signature dataset, $T=5026$. When $n=2$, a given start sequence is expected to be observed $~90.14$ and $~139.6$ times for the gestures and signatures respectively. When $n=3$, 
the expected numbers for gestures and signatures drop to $~2.50$ and $~3.88$. When $n=4$, they drop to $~0.07$ and $~0.11$. Since there are fewer than one observations for each start symbol when $n=4$, it is clear that models with this depth or greater will assign zero probability to almost all passwords and are thus unusable. Thus, we only consider n-gram models with $n=2$ and $n=3$.

\subsubsection{Smoothing}

Aside from assigning a small value for n to deal with uncovered passwords, a more common approach involves estimating the probabilities of the uncovered passwords using  smoothing methods. We tested two smoothing methods: 1) additive smoothing, and 2) Good-Turing smoothing~\cite{GoodTuring}. Additive smoothing adds a constant small value $\lambda$ to the counts of the Markov chain transition matrix. We assign $\lambda$  to the uncovered data, which are originally valued at zero, to make sure all theoretical possible data have some probability of occurring. Based on our tests, we selected $\lambda=0.01$. Good-Turing smoothing uses the observed total probability of class $r+1$ to estimate the total probability of class $r$. The total probability of class $r$ is a class of probability of transitions from a symbol to another symbol that has occurred r times in total.

\subsection{Optimizing Markov Chains for Recognition Passwords}

We have proposed SAX to represent recognition passwords with discrete symbols, and we have used Markov chains to estimate the probability of recognition passwords. We cannot judge a priori which configuration of Markov chains is most suitable for predicting recognition passwords. The optimal Markov chain should be most efficient at cracking recognition passwords. We will use guessing entropy~\cite{guessEntropy1,guessEntropy2} to demonstrate the cracking efficiency of the Markov chains.

\subsubsection{Guessing Entropy}
\label{sec:guessingentropy}

\begin{figure}[!t]
	\centering
	\includegraphics[width=1\columnwidth]{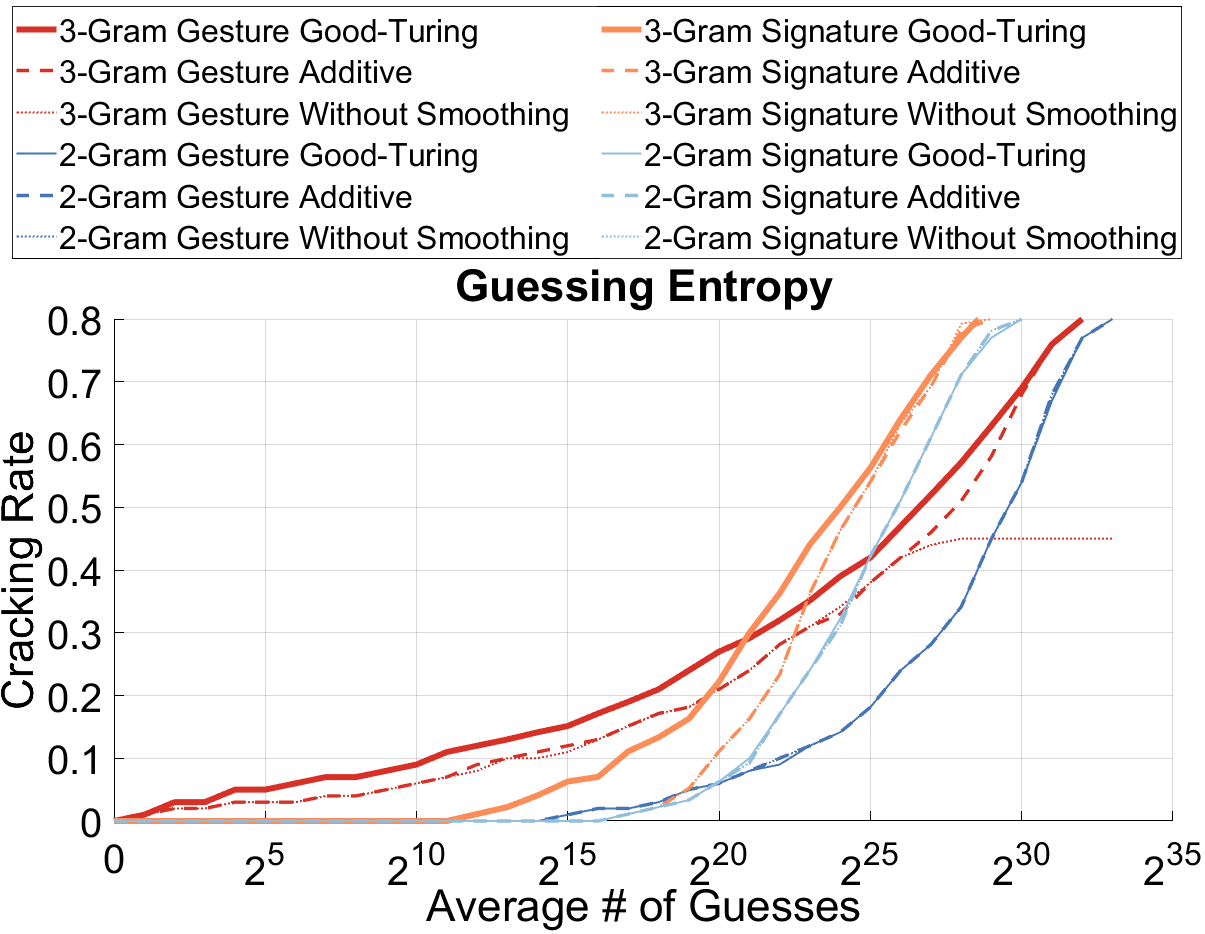}
	\caption{Guessing entropy of recognition passwords under six configurations of Markov chain: 
	2-gram Additive, 2-gram Good-Turing, 2-gram without smoothing, 3-gram Additive, 3-gram 
	Good-Turing, and 3-gram without smoothing. Generally, 3-gram Markov chains for gestures 
	and signatures are better than 2-gram models. There is no obvious difference between with 
	and without smoothing methods.}
	\label{fig:guessGesSig}
\end{figure}

Guessing entropy~\cite{guessEntropy1,guessEntropy2} measures the average number of guesses required to crack an entire set of passwords $X=\{x_1, x_2, ..., x_N\}$ in the optimal guessing order. Specifically, a guessing entropy curve represents the percentage of a dataset that is cracked as the number of guesses increases. It reflects the strength of the target passwords. Generally, a given guessing entropy means that it takes an average of so many guesses to crack some proportion of the data (see Figure~\ref{fig:guessGesSig}). An attacker ranks their guesses from most likely to least likely and proceeds in that order as they attack.

\subsubsection{Markov Chain Implementation}

In order to estimate the guessing entropy with the Markov chain, we performed a 10-fold cross validation. We first combined the three gesture datasets (FreeForm~\cite{free_form_gesture}, 
Wild~\cite{ESM}, and GuessAttack~\cite{guessingAttack}) into one dataset. Then, we split it into ten subsets with roughly the same number of accounts. For each training process, we selected one subset as a testing set and used the other nine subsets as a training set. We repeated this training process ten times.

\subsubsection{Comparison of Markov Chains}

We tested both 2-gram and 3-gram Markov chains with and without the two smoothing methods, Good-Turing and additive, for the two types of recognition passwords: gestures and signatures, as Figure~\ref{fig:guessGesSig} shows. We needed to find Markov models for both the upper- and lower-bound estimates for the security of passwords.

The 3-gram Markov chain with Good-Turing smoothing achieves the highest cracking rates with the same average number of guesses for both gesture and signature passwords. We found the 3-gram Markov chains to be more efficient than the 2-gram models. Taking gestures as an example, the 3-gram Markov chain with the Good-Turing smoothing method is at least ten percentage points higher in efficiency than the 2-gram Markov chain with the Good-Turing smoothing method. However, the choice of smoothing method does not have a significant impact on guessing entropy. For example, we found that the 2-gram Markov chains with additive and Good-Turing smoothing nearly overlap for both gestures and signatures. The difference between Markov chains with and without smoothing  is also not obvious with the exception of cases in which the target passwords have zero probabilities in the Markov chain. For example, we found that the guessing entropy of gestures based on a 3-gram model with and without Good-Turing smoothing are close to each other before $2^{26}$ guesses. Then, the guessing entropy of the Markov chain without smoothing is stable since the rest of the target passwords have zero probability under this model. Therefore, we selected the 3-gram Markov model with Good-Turing smoothing to model the upper bound of the password distribution and selected the 3-gram Markov model without smoothing to model the lower bound of the password distribution.

\section{Partial Guessing Metric}

Since guessing entropy, discussed above in Section~\ref{sec:guessingentropy}, is based on the cracking rate of a specific password dataset, the security of the passwords will be over-or-under-estimated depending on the quality of that dataset. 

Bonneau~\cite{partialGuess} proposed a partial guessing metric (or $\alpha-guesswork$) for user-chosen passwords based on the password distribution to overcome the problems inherent in guessing entropy. The partial guessing metric models a practical attack situation in which the attacker has knowledge of the general password distribution $\chi=\{x_1,x_2,\cdots\}$ with the goal of cracking a certain percentage of the passwords.

We define $\mu_{\alpha}(\chi)=min(j|\Sigma_{i=1}^j p(x_i)\geq \alpha)$ as the minimal number of 
needed guesses to crack a $\alpha$ proportion of passwords, and define 
$\lambda_{\mu_{\alpha}(\chi)}(\chi)=\lambda_{\mu_{\alpha}}(\chi)=\Sigma_{i=1}^{\mu_{\alpha}} 
p(x_i)$ as the the actual cracked proportion of passwords with $\mu_{\alpha}(\chi)$ guesses. 
Then, the partial guessing metric is defined as 

\begin{equation}
    G_{\alpha}(\chi)=(1-\lambda_{\mu_{\alpha}}) \cdot \mu_{\alpha} + \Sigma_{i=1}^{\mu_{\alpha}} p(x_i) \cdot i
\end{equation}

where the first term reflects a fraction of passwords that are not cracked within a given number of guesses while the second term reflects the minimum expected number of guesses needed to crack the fraction $\alpha$ of possible passwords selected by people. 

It is important to emphasize the key difference between guessing entropy and a partial guessing metric. Guessing entropy analyzes the cracking rate of a particular set of target passwords. In contrast, a partial guessing metric analyzes a fraction of the distributions of user-chosen passwords. For example, the estimation would vary when the set is non-representative of the population or if there is skew introduced by participants. Imagine a set with ten passwords, nine of which are cracked within $100$ guesses while the last one is cracked after $10^9$ guesses. This makes the guessing entropy higher than $10^8$, which is an extreme overestimation. The system is not secure since 90\% of the passwords were cracked within 100 guesses. This is why a partial guessing metric is a preferable security metric for comparing passwords\footnote{Partial guessing metric is \textit{not} used to adjust the 
parameters of the Markov chain. Guessing entropy is used to adjust the model because we 
measure performance by successfully cracking as many passwords in the training set as 
possible without consideration to how much work is required to crack those passwords. A Markov 
chain cannot be trained with partial guessing metric since it requires probabilities to be 
assigned to an entire distribution, whereas we need the Markov chain to assign those 
probabilities.}.

\section{Results}

We first present the major result that was made possible by this present work: a quantitative evaluation of the security of recognition passwords by the partial guessing metric. Then, we compare the partial guessing metric between recognition passwords and Android unlock patterns. Finally, we examine biases in the distribution of gesture passwords and signatures.

\subsection{Upper and Lower Bounds on Password Security}
\begin{figure}[!t]
\centering
\includegraphics[width=1\columnwidth]{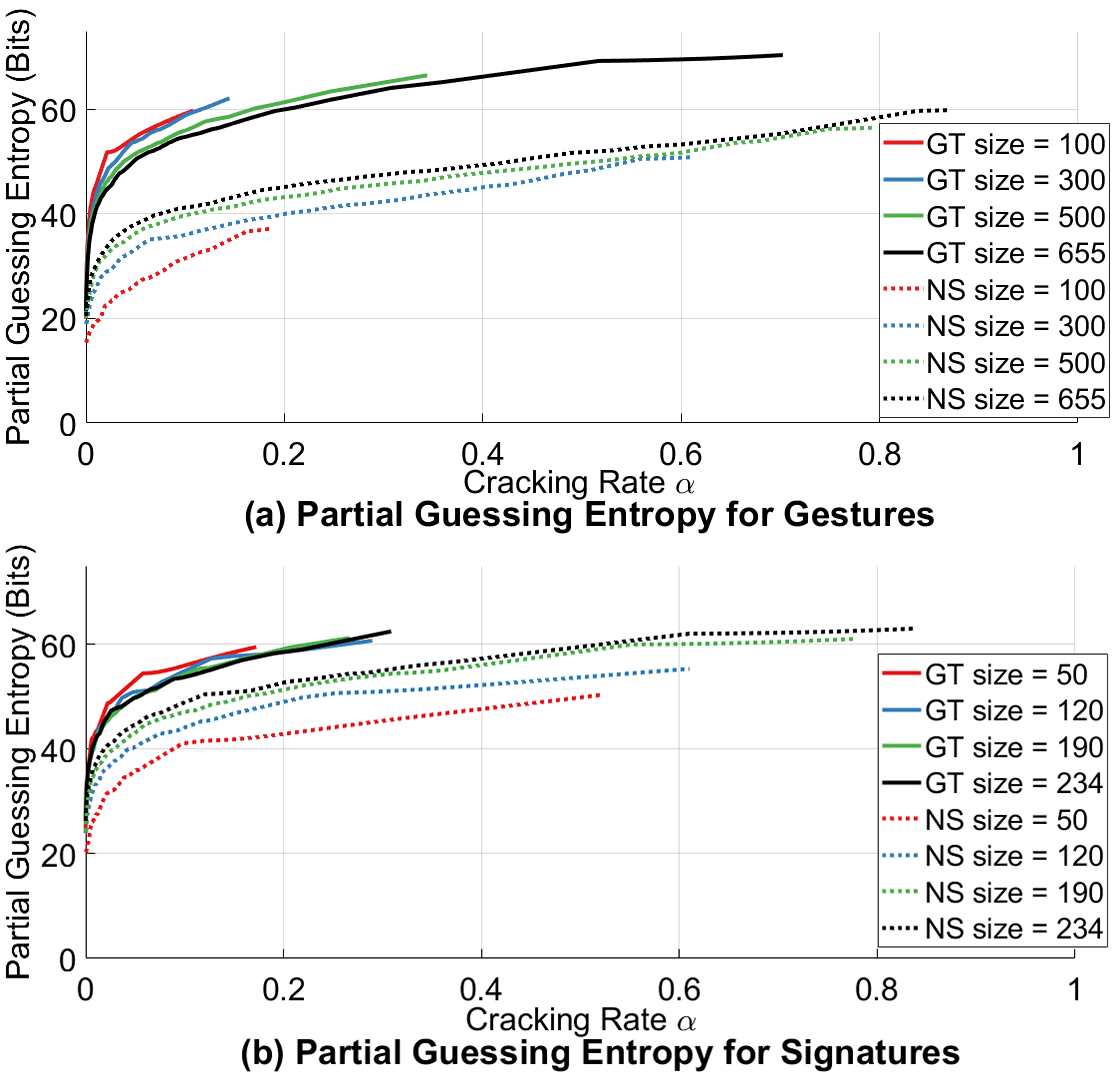}
\caption{Partial guessing metric of gestures and signatures that estimated based on different sizes of 
passwords datasets. We found the partial guessing metric curves with Good-Turing smoothing method decreases when the password datasets size increase. Similarly, we found the partial guessing metric curves without smoothing methods increases when the password datasets size increase.}
\label{fig:partialEntropyUpLowBounds}
\end{figure}

Figure~\ref{fig:partialEntropyUpLowBounds} shows that when the sizes of password datasets increase, the partial guessing metric estimation based on a Markov model with Good-Turing smoothing will decrease and the partial guessing metric based on a Markov model without smoothing will increase. This observation matches our analysis of the influence of the uncovered passwords on the password distribution, as Figure~\ref{fig:passwordDistributionSmooth} shows. 

Based on the observations of Figure~\ref{fig:partialEntropyUpLowBounds}, if we keep increasing the size of the datasets, the partial guessing metric curve of Good-Turing method will keep decreasing and the curve without smoothing method will keep increasing. Eventually, they will converge to the partial guessing metric of the actual password distribution.

In conclusion, with the current size of our recognition password dataset, we are able to provide upper and lower bounds on the partial guessing metric of gestures and signatures.

\subsection{Comparison to Android Unlock Patterns}

\begin{figure}[!t]
\centering
\includegraphics[width=1\columnwidth]{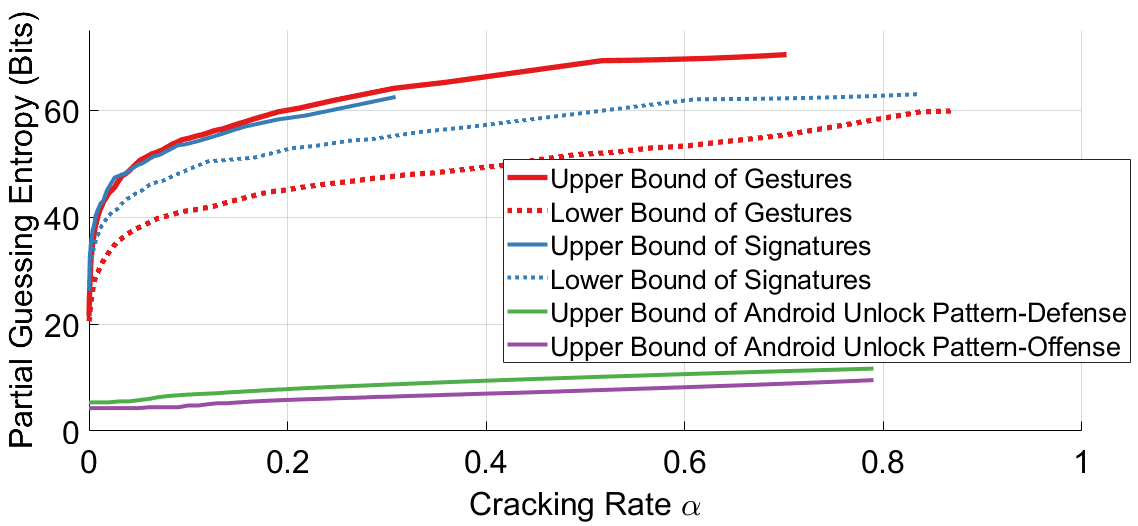}
\caption{Partial guessing metric for different types of passwords. We found the lower bounds of 
security of recognition passwords (gestures and signatures) are considerably higher than Android 
unlock patterns.}
\label{fig:partialEntropyPattern}
\end{figure}

Figure~\ref{fig:partialEntropyPattern} shows that recognition passwords, gestures, and signatures have a higher partial guessing metric than Android unlock patterns. For gestures, the lower bound of the partial guessing metric is 45 bits when the cracking rate $\alpha = 0.2$. This is 37 bits higher than the upper bound of defense-oriented Android unlock patterns with $\alpha = 0.2$. The password distribution of Android unlock patterns is modeled on the 3-gram Markov model with additive smoothing~\cite{android_pattern}. Similarly, for signatures, the lower bound of the partial guessing metric is 52 bits when the cracking rate $\alpha = 0.2$, which is also much higher than the corresponding metric for Android unlock patterns.

\subsection{Human Bias in Recognition Passwords}

\begin{figure}[!t]
\centering
\includegraphics[width=1\columnwidth]{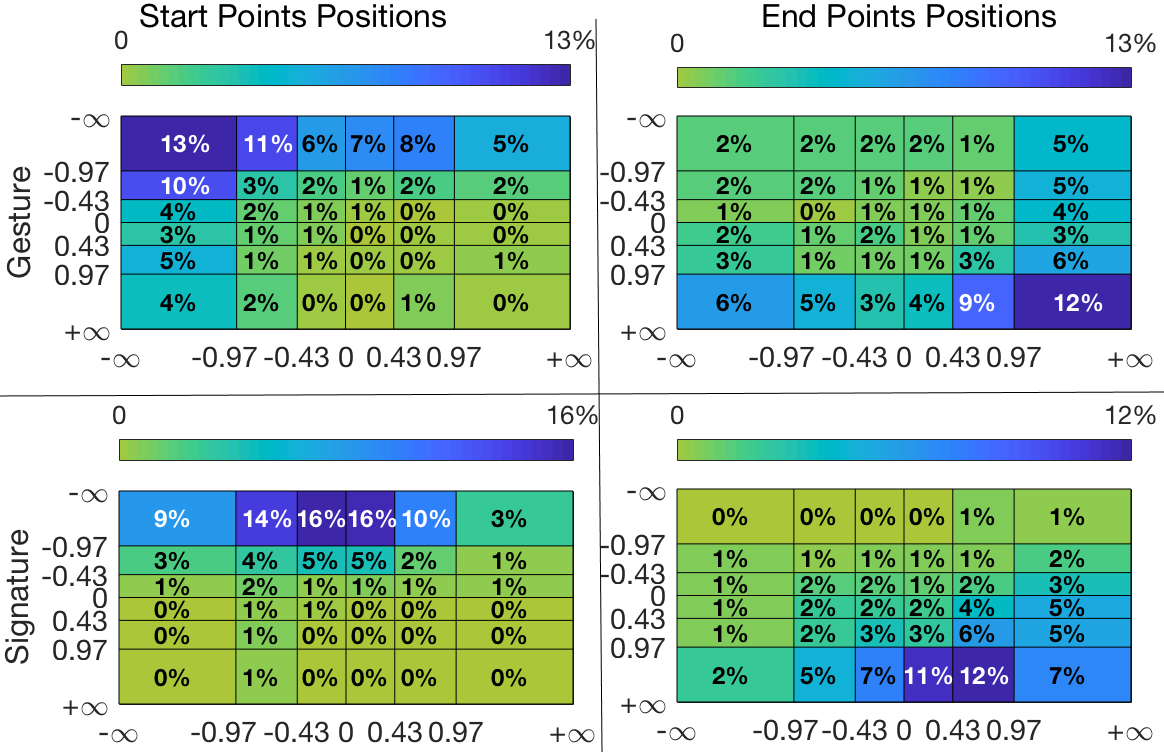}
\caption{ Biases of the start and end points of gestures (upper two figures) and signatures (lower two figures). Generally, for both gestures and signatures, the passwords start from the top left corner of the screen and finish at the bottom right corner.}
\label{fig:biasSigGes}
\end{figure}

We studied human bias in gesture and signature passwords. Specifically, we analyzed the distribution of the password starting and ending points and the connecting pattern in the passwords. Figure~\ref{fig:biasSigGes} shows where the starting and ending points of gestures and signatures are found on a screen across the datasets. We found that passwords usually start from the top left corner and end at the bottom right corner of the screen. A possible reason for this is that writing habits in Western cultures move from left to right and from top to bottom.

We examined the covered percentages of 2-gram and 3-gram variations. However, we did not find obvious bias in those two models that we might suggest users avoid. We include them as Figure~\ref{fig:biasNgram} in Appendix A for reference.

\section{Discussion and Conclusions}

\textbf{We have presented the first paper to quantitatively evaluate the security of recognition passwords using a partial guessing metric. } We were able to do this because we are the first to estimate the distribution of recognition passwords using Markov chains. Specifically, we have demonstrated that recognition passwords given as time-series data can be represented as symbolic expressions using SAX. The symbolic alphabet can be combined with datasets of user-chosen passwords to compute Markov chains for a given method. These Markov chains can then be used to approximate the full password distribution by estimating the likelihood of unknown passwords. With the estimated password distribution, we quantitatively evaluated the security of recognition passwords by calculating their partial guessing metrics.

\textbf{We estimated the security of recognition passwords using upper and lower bounds with a relatively small dataset.} For novel passwords, there are always concerns that the dataset is not large enough to reflect the full password distribution. However, it is still necessary to estimate the security of novel passwords before they are widely employed. As Figure~\ref{fig:passwordDistributionSmooth} shows, the two strategies: (1) assigning zero probability (2) assigning a small probability to all of the uncovered passwords will lead to the upper bound and the lower bound on the password distribution estimation. To verify our analysis on the influence of password security estimation by these two methods, we sub-sampled the recognition password datasets to various smaller sizes. Figure~\ref{fig:partialEntropyUpLowBounds} shows that the changing trends of partial guessing metrics with different password dataset sizes match our analysis. These two bounds of recognition passwords allow a direct, numeric comparison to the security of matching passwords as well as any other recognition password.

\textbf{We quantitatively compared the security of recognition passwords to that of Android unlock patterns.} We have quantitatively shown that the security of gesture and signature passwords has a higher partial guessing metric than the Android pattern unlock method~\cite{free_form_gesture,back_authen,guessingAttack}. Prior work made arguments about security in three different ways: (1) quantifying the amount of expressive information contained in free-form gestures~\cite{free_form_gesture}, (2) calculating the size of the total password space~\cite{free_form_gesture}, and (3) segmenting the grid into patterns and calculating random entropy. However, these security measures are not as reliable as a partial guessing metric because they do not address how users or attackers behave~\cite{partialGuess}. The higher the partial guessing metric, the more secure the password system is. After assessing the number of guesses per account an attacker needs to deploy in order to crack an $\alpha$-sized portion of all accounts on multiple different scales (Figure~\ref{fig:guessGesSig}), we have shown with a rigorous approach and direct comparison that the lower bounds of gesture and signature passwords have a higher partial guessing metric than the upper bounds of Android unlock patterns.

\textbf{We used distinguishability between recognition passwords to confirm the validity of discretizing the recognition passwords by SAX.} A reasonable concern that arises in the discretization of recognition passwords asks whether the newly discretized passwords can represent original passwords. As an authentication method, the primary purpose of the recognition password feature is to distinguish various users' passwords. In other words, we had to keep both false positive errors and false negative errors low. Therefore, the distinguishability of passwords can be measured by ROC and AUROC. To show that SAX does not hurt the distinguishability of recognition passwords, we implemented  state-of-the-art recognition methods for passwords and examined their ROC and AUROC. We found that SAX achieves ROC and AUROC results comparable to other recognition methods.

\textbf{Limitations.}
Two of the gesture datasets (FreeForm~\cite{free_form_gesture} and 
GuessAttack~\cite{guessingAttack}) and all of the signature datasets~\cite{susig,mcyt,svc2004} in our study were collected in the laboratory, which in principle could affect the participants' selection of gestures and signatures as passwords~\cite{ESM}. By comparing the gestures collected from the laboratory~\cite{free_form_gesture,Policy,guessingAttack} and in the wild~\cite{ESM}, we do not find any evidence of a clear difference between the passwords generated under these two types of environments. Thus, we cannot estimate the influence of experimental environments on our results.

The three gesture datasets were collected across different studies, and we aggregated them into one gesture dataset. Since all of the participants in the three datasets were asked to create gestures as passwords, different experimental setups will enlarge the diversity and coverage of the gesture passwords. Aggregating datasets from different studies does not damage our analysis of the password distribution.

\textbf{Baseline and Dataset Size.} There are natural concerns about dataset size when it comes to estimating the partial guessing metric. The upper and lower bounds placed on the partial guessing metric solved this question with fixed parameter values of SAX. However, a larger size for the dataset may also increase the parameter values of SAX. Generally, a larger dataset size may (1) increase the total size of the symbolic alphabet used in SAX representation, and (2) assign probabilities to the newly enumerated passwords.

\uline{Alphabet Size}. The size of the alphabet provides a baseline for the security result. The size of the password space increases as the alphabet size increases since passwords that are clustered together become separated out more easily. This reduces the overall size of the weak set; see Figure~\ref{fig:Figure_recognizerLowerBound} for an example. Additionally, Table 3 in Appendix B provides numeric evidence that the partial guessing metric increases as the alphabet size increases. Collecting additional passwords would not decrease the overall size of the alphabet. A larger number of passwords would require a larger alphabet to represent it if there are new attributes that need to be accounted for (see Figure~\ref{fig:Figure_SAX} for a depiction of how the alphabet maps to a password). As such, if we collect many more passwords, we expect the size of the estimated password space to increase since the alphabet size will increase. This will increase both the upper and the lower bounds of the security estimation by the partial guessing metric.

\uline{Password Probabilities}. The security of passwords is related to the cracking effectiveness of our Markov chain, which attacks passwords ordered by probability from high to low. The password probability calculated by the Markov chain is mainly affected by the weak sets of gestures and signatures used to train it. These weak sets, as in text passwords~\cite{partialGuess}, are concentrated and significantly smaller than the theoretical password space. If the data sample is representative, then collecting more user-generated passwords is not expected to change the partial guessing metric because it would not significantly affect the ratio of the size of the weak set to the entire password space. Therefore, additional passwords in the dataset might stretch the tails of the estimated passwords distribution -- more unlikely passwords would be sampled if the sample size were increased -- but this would not significantly impact the ratio of the weak set to the whole space because those points are the most discoverable by definition. Because we sampled the general distribution repeatedly, we have an estimate of the true distribution that includes the weak set. Because the partial guessing metric is functionally dependent on the coverage of the weak set, the stability of the weak set size means that the guessing metric will not be impacted by the tails of the distribution.

\uline{Distributional Bias}. If the collected data is not representative and is in fact oversampling either a subset of the weak set or the tails of the distribution, then this would give a misleading estimate of the partial guessing metric. This would happen because the data would misinform the Markov chain about the probability of the enumerated passwords from the SAX representation. We have evidence that this is not likely to have happened. Liu et al.~\cite{guessingAttack} examined the relative frequencies of different gesture password categories across several gesture datasets in their cracking paper and found that the relative frequencies of groups (shapes, words, symbols) are equal despite being collected at different times by different groups (see Table 2 in Liu et al.~\cite{guessingAttack}). This is evidence that the collected gestures are being sampled from a general distribution and not from the tails of such a distribution.

\uline{Summary.} Our core contribution is true irrespective of the total set of data one might possibly collect on user-generated passwords: we have established a baseline for estimating the upper and lower bounds of the partial guessing metrics of recognition passwords. While password enumeration and the Markov chain's performance can only be improved with future data, our current estimates are the best given all publicly available data at this time. More data can only improve a model, not negate the utility of the existing model.

\section*{Acknowledgments}

This material is based upon work supported by the National Science Foundation under Grant Number 1750987. Any opinions, findings, and conclusions or recommendations expressed
in this material are those of the authors and
do not necessarily reflect the views of the National
Science Foundation. Additional material available at
\url{http://scienceofsecurity.science}.

\balance{}

{\footnotesize \bibliographystyle{acm}
\bibliography{SAX_SPversion}}

\newpage
\begin{appendices}

\subsection*{Appendix}
\section{Human Bias in N-Gram Markov Chain}
\begin{figure}[!t]
\centering
\includegraphics[width=1\columnwidth]{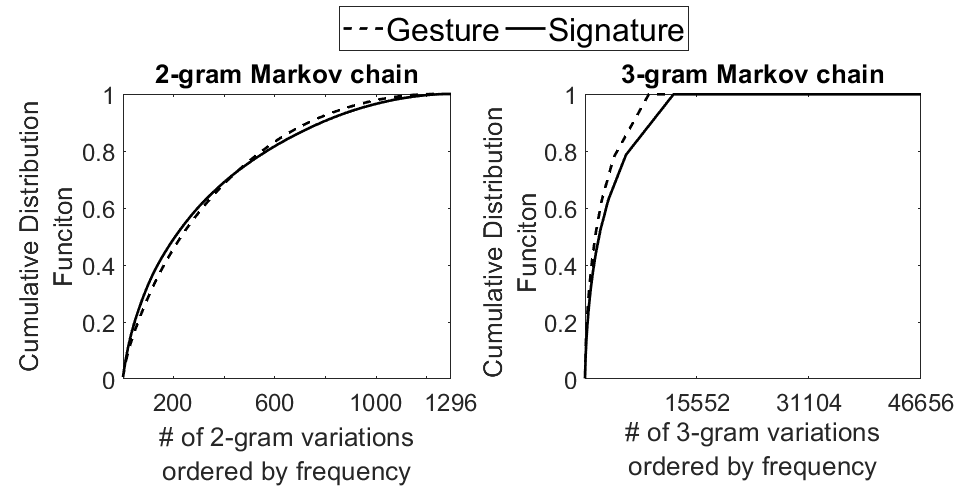}
\caption{Coverage of the different 2-gram and 3-gram variations. There are totally 1296 variations in the 2-gram model and 46656 variations in the 3-gram model. We found some of the variations are more likely to be selected by users. But the bias in selections is not obvious.}
\label{fig:biasNgram}
\end{figure}

The left and right figures in Figure~\ref{fig:biasNgram} shows the distributions of the 2-gram model and 3-gram model, respectively. We found no obvious bias in the two models. For example, the top 200 frequent 3-grams only cover $13.87\%$ of all the variations in signature passwords. 

\section{Parameter Selection Effect on Security}

Table~\ref{tab:AlphaN} shows the partial guessing entropy of different cracking rates ($\alpha$=0.1, 0.2, 0.5) for gestures and signatures of Good-Turing 3-gram Markov chains with different SAX parameter selections ($\beta$=4, 5, 6, 7, 8; $\omega$ = 4, 5, 6, 7, 8). In order to show the effect on partial guessing metric estimation with smaller parameters in our method, we mainly select the values of $\alpha$ and $\omega$ that are less than the optimal value pair, $\beta = 6$ and $\omega = 8$, since the larger parameter values always lead to a higher estimation of partial guessing metric. As a reminder, $\beta$ represents the size of alphabet of symbols for 1-D SAX. Since the gesture and signature are discretized by 2-D SAX, the size of alphabet should be $\beta^2$. For example, the size of the theoretical full passwords space with $\beta = 4$ and $\omega = 4$ is $(\beta^2)^{\omega}=65536$.

We found the selection of SAX parameters has large effects on the security estimation of gestures and signatures.  With the decrease of $\beta$ and $\omega$, the estimated security of gestures and signatures also decrease. As a reminder, we selected the optimal pair of $\beta=6$ and $\omega=8$ based on AUROC values in Figure~\ref{fig:Figure_AlphaOmega}, which measures the distinguishability between positive and negative samples in authentication system. The decrease of   $\beta$ and $\omega$ will also lead to the decrease of distinguishability in gestures and signatures passwords and significantly harm the recognition performance of those passwords.

\begin{table}[!t]
\footnotesize
\center

\subfloat[$\alpha=0.1$]{\begin{tabular}{cp{0.33 cm}p{0.33 cm}p{0.33 cm}p{0.33 cm}p{0.4 cm}|p{0.33 cm}p{0.33 cm}p{0.33 cm}p{0.33 cm}p{0.33 cm}}
\hline
\multicolumn{1}{l}{}   & \multicolumn{3}{l}{Gesture} & \multicolumn{1}{l}{$\omega$} & \multicolumn{1}{l|}{} & \multicolumn{3}{l}{Signature} & \multicolumn{1}{l}{$\omega$} & \multicolumn{1}{l}{} \\ \hline
\multicolumn{1}{c|}{$\beta$} & 4      & 5       & 6       & 7                     & 8                     & 4        & 5        & 6       & 7                     & 8                    \\ \hline
\multicolumn{1}{c|}{4} & 12.9    & 19.2    & 25.7    & 31.1                  & 36.7                  & 11.8     & 17.4     & 23.4    & 30.9                  & 37.2                 \\
\multicolumn{1}{c|}{5} & 16.2    & 24.1    & 31.6    & 38.5                  & 45.3                  & 13.5     & 20.7     & 29.4    & 36.9                  & 45.3                 \\
\multicolumn{1}{c|}{6} & 20.6    & 29.8    & 38.5    & 46.3                  & 56.4                  & 15.9     & 25.5     & 34.3    & 43.3                  & 54.8                 \\
\multicolumn{1}{c|}{7} & 24.5    & 33.1    & 44.4    & 53.5                  & 63.9                  & 18.3     & 26.1     & 39.0    & 48.9                  & 60.2                 \\
\multicolumn{1}{c|}{8} & 27.9    & 39.4    & 50.3    & 61.4                  & 71.2                  & 21.6     & 33.6     & 45.0    & 55.5                  & 65.3                 \\ \hline
\end{tabular}}
\qquad
\subfloat[$\alpha=0.2$]{\begin{tabular}{cp{0.33 cm}p{0.33 cm}p{0.33 cm}p{0.33 cm}p{0.4 cm}|p{0.33 cm}p{0.33 cm}p{0.33 cm}p{0.33 cm}p{0.33 cm}}
\hline
\multicolumn{1}{l}{}   & \multicolumn{3}{l}{Gesture} & \multicolumn{1}{l}{$\omega$} & \multicolumn{1}{l|}{} & \multicolumn{3}{l}{Signature} & \multicolumn{1}{l}{$\omega$} & \multicolumn{1}{l}{} \\ \hline
\multicolumn{1}{c|}{$\beta$} & 4       & 5       & 6       & 7                     & 8                     & 4        & 5        & 6       & 7                     & 8                    \\ \hline
\multicolumn{1}{c|}{4} & 15.1   & 21.5 	& 28.2 		& 34.0                 & 40.0                  & 14.2    &  19.8    &  26.3     &  33.8              &   41.2                 \\
\multicolumn{1}{c|}{5} & 18.9     & 27.2    & 35.6     & 42.4                 & 49.7                 & 15.9       & 24.1    & 32.2      & 40.9              & 50.5                \\
\multicolumn{1}{c|}{6} & 23.9   & 33.8    & 42.4      & 50.4                 & 60.6                  & 19.5      & 28.9    &   37.8     &  47.5             &   59.3                \\
\multicolumn{1}{c|}{7} & 27.3    & 37.2    & 48.3      & 57.8                 & 68.9                  & 21.9       &  29.1   &   43.9     &   52.8           &    65.3                  \\
\multicolumn{1}{c|}{8} & 31.0    & 42.4    & 54.0       & 65.5                 &  75.2                 & 26.7      & 38.5   & 49.8       & 61.0             &     69.4                \\ \hline
\end{tabular}}
\qquad
\subfloat[$\alpha=0.5$]{\begin{tabular}{cp{0.33 cm}p{0.33 cm}p{0.33 cm}p{0.33 cm}p{0.4 cm}|p{0.33 cm}p{0.33 cm}p{0.33 cm}p{0.33 cm}p{0.33 cm}}
\hline
\multicolumn{1}{l}{}   & \multicolumn{3}{l}{Gesture} & \multicolumn{1}{l}{$\omega$} & \multicolumn{1}{l|}{} & \multicolumn{3}{l}{Signature} & \multicolumn{1}{l}{$\omega$} & \multicolumn{1}{l}{} \\ \hline
\multicolumn{1}{c|}{$\beta$} & 4       & 5       & 6       & 7                     & 8                     & 4        & 5        & 6       & 7                     & 8                    \\ \hline
\multicolumn{1}{c|}{4} & 18.9    & 25.9     & 32.9      & 40.0                 & 46.6                 & 17.9      &  25.2    &  33.0   &  40.0              &  47.2                  \\
\multicolumn{1}{c|}{5} & 24.5    & 32.5      & 41.1      & 48.5                 & 56.9                  & 23.3    & 30.0     & 39.5     & 47.4              & 56.2                 \\
\multicolumn{1}{c|}{6} & 29.0     & 38.5     & 46.3    & 56.8                 & 67.7                   & 27.2     &  36.5    &  44.9     &  54.9             &   65.8             \\
\multicolumn{1}{c|}{7} & 32.5      & 42.5     & 53.9    & 64.0                 & 74.0                  & 30.3     &   35.3    &   51.4     &   61.1            &     70.5                  \\
\multicolumn{1}{c|}{8} & 36.3     & 47.8      & 58.9     & 70.5                &   78.2                 & 33.6    & 45.9      & 57.1         & 68.0            &   73.3                 \\ \hline
\end{tabular}}
\caption{Comparing partial entropy with different cracking rates ($\alpha$=0.1, 0.2, 0.5) with different parameters values ($\omega$ = 4, 5, 6, 7, 8; $\beta$ = 4, 5, 6, 7, 8) for gestures and signatures. The decrease of $\omega$ and $\beta$ leads to lower estimated security for gestures and signatures. However, since the decrease of $\omega$ and $\beta$ also lead to worse recognition performance as shown in Figure~\ref{fig:Figure_AlphaOmega}, we need to select the values of $\omega$ and $\beta$ that optimizes recognition performance.}
\label{tab:AlphaN}
\end{table}

\end{appendices}
\end{document}